\documentclass[12pt,a4paper,english]{article}
\usepackage{titling}
\usepackage[affil-it]{authblk}
\usepackage{longtable}
\usepackage{lscape}
\usepackage{graphicx}
\usepackage{epsfig}
\usepackage{dsfont}
\usepackage{amsfonts}
\usepackage{amsmath}
\usepackage{amsthm}
\usepackage{amssymb}
\usepackage[english]{babel}
\usepackage[utf8]{inputenc}
\usepackage{mathrsfs}
\usepackage{pdfpages}
\usepackage{tabu}
\usepackage{simplewick}
\usepackage[hyperindex,breaklinks,colorlinks,linkcolor=blue,citecolor=blue,urlcolor=blue]{hyperref}
\usepackage{bm}
\usepackage{multirow}
\usepackage{hhline}
\usepackage[normalem]{ulem}
\usepackage{cite}
\usepackage{xcolor}
\usepackage{fancyhdr}
\usepackage[framemethod=TikZ]{mdframed}
\usepackage{soul}
\setuldepth{Obtain}
\usepackage{pgfgantt}
\usepackage{subcaption}
\usepackage[left=2.6cm,right=2.6cm,top=2.55cm,bottom=2.6cm]{geometry}

\usepackage{authblk}
\usepackage{lineno}

\makeatletter
\g@addto@macro\bfseries{\boldmath}
\makeatother

\definecolor{ForestGreen}{RGB}{34,139,34}

\newcommand{\D}[1]{{\rm d}}
\newcommand{\Br}{{\mathcal B}}

\usepackage[many]{tcolorbox} 
\definecolor{main}{HTML}{5989cf} 
\newtcolorbox{mybox}{
    boxrule = 1.5pt,
    colframe = main,
    rounded corners,
    arc = 5pt   
}

\title{Kaon Physics: A Cornerstone for Future Discoveries}

 \author[1]{Jason Aebischer}
 \author[2]{Atakan Tugberk Akmete}
 \author[3]{Riccardo Aliberti}
 \author[4,5]{Wolfgang Altmannshofer}
 \author[6,7,8]{Fabio Ambrosino}
 \author[9]{Roberto Ammendola}
 \author[10]{Antonella Antonelli}
 \author[11,12]{Giuseppina Anzivino}
 \author[13,14]{Saiyad Ashanujjaman}
 \author[15]{Laura Bandiera}
% ...
\affil[1]{CERN, Espl. des Particules 1, 1217 Gen\`{e}ve, Switzerland}
\affil[2]{Institut f\"ur Physik and PRISMA$^+$  Cluster of Excellence, Johannes Gutenberg Universit\"at,  55099 Mainz, Germany}
\affil[3]{Institut f\"ur Kernphysik, Johannes Gutenberg Universit\"at,  55099 Mainz, Germany}
\affil[4]{Department of Physics, University of California Santa Cruz, Santa Cruz, CA 95064, USA}
\affil[5]{Santa Cruz Institute for Particle Physics, University of California Santa Cruz, Santa Cruz, CA 95064, USA}
\affil[6]{Dipartimento di Fisica "Ettore Pancini", Università degli Studi di Napoli Federico II, Via Cintia 80126 Napoli, Italy}
\affil[7]{INFN Sezione di Napoli, Napoli, Italy}
\affil[8]{Scuola Superiore Meridionale, Via Mezzocannone 4, 80134 Napoli, Italy}
\affil[9]{INFN Sezione di Roma Tor Vergata, Via della Ricerca Scientifica 1, 00133 Roma Italy}
\affil[10]{INFN Laboratori Nazionali di Frascati, 00044 Frascati RM, Italy}
\affil[11]{Department of Physics and Geology, University of Perugia, 06123 Perugia, Italy}
\affil[12]{INFN Sezione di Perugia, Perugia, Italy}
\affil[13]{Institute for Theoretical Particle Physics (TTP), Karlsruhe Institute of Technology, 76131 Karlsruhe, Germany}
\affil[14]{Institute for Astroparticle Physics (IAP), Karlsruhe Institute of Technology, Hermann-von-Helmholtz-Platz 1, D-76344 Eggenstein-Leopoldshafen, Germany}
\affil[15]{INFN Sezione di Ferrara, Via Saragat 1, 44122 Ferrara, Italy}
% ...

 \usepackage{hep-title}
 \preprint{CERN-TH-2025-066}

\begin{document}

\maketitle

\thispagestyle{empty}

\begin{abstract}
The kaon physics programme, long heralded as a cutting-edge frontier by the European Strategy for Particle Physics, continues to stand at the intersection of discovery and innovation in high-energy physics (HEP). With its unparalleled capacity to explore new physics at the multi-TeV scale, kaon research is poised to unveil phenomena that could reshape our understanding of the Universe. 
This document highlights the compelling physics case, with emphasis on exciting new opportunities for advancing kaon physics not only in Europe but also on a global stage. As an important player in the future of HEP, the kaon programme  promises to drive transformative breakthroughs, inviting exploration at the forefront of scientific discovery.
\end{abstract}

\flushbottom

\newpage

\setcounter{page}{1}

\section{Introduction}
The electroweak hierarchy problem and the unexplained hierarchies of the Yukawa couplings of the fermions, the so-called flavour problem, are among the most puzzling features of the Standard Model (SM), indicating that a more fundamental theory is waiting to be discovered, which could shed light on these questions.  Rare kaon decays provide a very sensitive and, most importantly, unique probe of such extensions.

{As with all indirect probes of heavy new physics (NP), the primary goal is to identify processes that are both highly suppressed and precisely predictable within the SM, while being exceptionally sensitive to physics beyond the SM. The 
$K\to\pi\nu\bar\nu$
decay modes, along with specific observables in rare $K_{L,S}$	
  decays (discussed below), meet these criteria. They are  the only available probes of flavour-changing neutral-current (FCNC) transitions of the type $s\to d \nu\bar\nu$ and $s\to d \ell \bar\ell$  ($\ell=e,\mu$), which are significantly suppressed within the SM framework and exhibit heightened sensitivity to NP effects. Amongst these, there are gold-plated channels, which  are largely unaffected by long-distance dynamics, enabling their SM prediction to achieve remarkable theoretical precision.}

In any general extension of the SM that addresses the hierarchy  and/or the flavour problem, these transitions might not be suppressed as in the SM. They are, therefore, ideally suited for the study of violation of low-energy accidental flavour symmetries and obtaining clues on possible directions for Beyond Standard Model (BSM) physics. 
The $\Br(K^+ \to \pi^+ \nu\bar \nu)$ and $\Br(K_L \to \pi^0 \nu\bar  \nu)$ decays are conceptually similar to theoretically clean and precise weak-scale observables, such as the $W$ mass or the Higgs self-coupling. Similarly to those, rare $K$ decays probe electroweak dynamics. 
However, rare $K$ decays probe it in a different and less tested sector connected to the flavour problem.
Not surprisingly, there are many examples of well motivated BSM scenarios in the literature, fully consistent with present high-energy data, which would give rise to large deviations from the SM in both decay modes, as can also be seen in model-independent studies (see, e.g., Refs.~\cite{Straub:2013zca, Buras:2015yca, Ishiwata:2015cga, Bordone:2017lsy, Fajfer:2018bfj, Mandal:2019gff, Marzocca:2021miv, Crosas:2022quq, DAmbrosio:2023irq, DAmbrosio:2024rxv}).
A further unique aspect of  $K\to\pi\nu\bar\nu$ decays is that they are sensitive to the interaction of light quarks ($s$ and $d$) with third-generation leptons (i.e., $\nu_\tau$).
This unique feature enhances their sensitivity to motivated BSM scenarios shedding light on the origin of the flavour hierarchies (see, e.g., Refs.~\cite{Bordone:2017lsy,Crosas:2022quq,Davighi:2023iks}).
In addition to being well motivated, such models can also naturally connect to hints of excesses in the present data over the SM predictions in heavy-flavour transitions, such as $b\to c\tau\nu$ and $b\to s\nu\bar\nu$ transitions (see, e.g., Ref.~\cite{Allwicher:2024ncl}), as measured at the LHCb and Belle(II) experiments.

Loosely speaking, the motivated BSM theories that can be tested via rare $K$ decays fall into the same categories as those searched for at HL-LHC and, in the future, at FCC-ee, i.e., theories with new heavy particles in the TeV to tens of TeV regime. 
Limits on specific exotic particles, or new contact interactions, from rare $K$ decays can even exceed 100~TeV~\cite{Buras:2014zga}.  
Such strong limits apply to new states that violate some of the approximate accidental symmetries of the SM, lifting most of the suppressions of FCNCs among the first two generations.
{\bf These high mass-scale sensitivities are manifestations of the uniqueness of rare $K$ decays in probing the flavour structure of physics beyond the SM, in ways that are not possible to test at HL-LHC nor at FCC.}

For many years, experimental studies of kaon decays have played a unique role in propelling the development of the SM. As in other branches of flavour physics, the continuing experimental interest in the kaon sector derives from the possibility of conducting precision measurements, particularly of suppressed or rare processes, which may reveal the effects of NP. Because of the relatively small number of kaon decay modes and the relatively simple final states, combined with the relative ease of producing intense kaon beams, kaon decay experiments are in many ways the quintessential intensity-frontier experiments.

The NA62 experiment at CERN is dedicated to studies of $K^+$ decays and is foreseen to conclude in 2026. After achieving in 2024 the first observation at $5\sigma$ of the ultra-rare decay $K^+ \rightarrow \pi^+ \nu \bar\nu$ with a partial dataset, the entire NA62 dataset will allow one to reach 15\% precision in the measurement of its branching ratio. The decay-in-flight technique successfully established by NA62, when combined with recent detector and DAQ advancements, would allow to build a follow-up experiment to reach 5\% precision, matching the theory uncertainty.
After 2026, the KOTO experiment at J-PARC will remain the only experiment worldwide dedicated to study rare kaon decays. With its entire dataset, KOTO will reach the SM sensitivity for the branching ratio of the ultra-rare decay $K_L \rightarrow \pi^0 \nu \nu$ by its conclusion in this decade. A much-upgraded experiment, KOTO~II, is planned, still at J-PARC, that can reach a $5\sigma$ observation for the first time. Thanks to further modifications of the setup, measurements of other rare $K_L$ decays, most notably $K_L \rightarrow \pi^0 \ell^+ \ell^-$, will also become possible for the first time.

Detection of ultra-rare kaon decays is extremely challenging experimentally and therefore requires a technology advancement in multiple directions. Dedicated experiments like NA62 that are of a scale smaller than for example LHC experiments, allow one to push detector technology that later could serve larger enterprises, with what can be seen as prototypes of innovation. Examples were the Gigatracker and the STRAW tracker detectors at NA62. The next-generation kaon experiments could serve as testing ground for the technologies needed for FCC experiments.

The rest of this document focuses on the channels which are the cleanest and most sensitive to BSM and that can potentially be probed experimentally in the next decades. Sections 2 and 3 are about FCNC $K^+$ and $K_L$ decays, respectively, while Section 4 is about $K_S - K_L$ interference. These sections have a theoretical part, summarising our theoretical understanding within the SM,  and an experimental part where experimental prospects are provided. Section 5 highlights the specific potential for SM tests and BSM searches. The document then concludes with the main recommendations.

\section{FCNC $K^+$ decays}
\subsection{Theory}
The current SM prediction for the branching ratio of $K^+ \to \pi^+ \nu \bar{\nu}$ semileptonic rare decay is 
$\mathcal{B}(K^+ \to \pi^+ \nu \bar{\nu}) = 8.38(17)(25)(40) \times 10^{-11}$ \cite{Brod:2021hsj,DAmbrosio:2022kvb,Aebischer:2022vky,Anzivino:2023bhp},
where the uncertainties correspond to short-distance contributions \cite{Buchalla:1998ba,Misiak:1999yg,Buras:2006gb,Brod:2008ss,Brod:2010hi}, long-distance contributions \cite{Isidori:2005xm,Lunghi:2024sjy}, and parametric inputs, respectively. Hence, the dominant uncertainty is parametric, primarily arising from CKM inputs, especially from the value for $|V_{cb}|$. Significant improvements are expected in the next 5-10 years, both on the extraction of the CKM elements~\cite{FlavourLatticeAveragingGroupFLAG:2024oxs} and on first-principle predictions for long-distance effects from lattice QCD simulations \cite{Christ:2016eae,Bai:2017fkh,Bai:2018hqu,Christ:2019dxu,Anzivino:2023bhp}. This will reduce the current total uncertainty of $\sim 6\%$ to $\sim 2\%$, which is, as illustrated in the next section, well below the precision of designs for next-generation experiments. 
This process also provides a rare opportunity to compare and test different theoretical methods used to predict the branching fraction of $K^+ \to \pi^+ \nu \bar{\nu}$. Such methods comprise Lattice QCD~\cite{Isidori:2005tv,Christ:2015aha,Christ:2016mmq,RBC:2022ddw}, chiral perturbation theory, and dispersive analysis. The interplay of these techniques offers a unique way to deepen our understanding of non-perturbative QCD effects and assess the robustness of current theoretical methods.

Given this excellent theoretical status and prospects, the $K^+ \to \pi^+ \nu \bar{\nu}$ decay probes not only electroweak interactions but also fundamental aspects of the flavour problem in the weakly explored $s \to d$ transitions. The very suppressed prediction in the SM enhances the $K^+ \to \pi^+ \nu \bar{\nu}$  sensitivity to possible contributions from NP, {allowing it to probe energy scales ranging from a few TeV up to hundreds of TeV, on par with or even superior to current energy-frontier collider experiments, or heavy-flavour experiments at the intensity frontier~\cite{Buras:2014zga,Isidori:2010kg}.} 
In fact, in high-energy experiments at $pp$ colliders, it is difficult to tag the strange jets. At future colliders like FCC-ee it would be possible to access flavour-changing $s\to d$ couplings, but never in combination with neutrinos. In general, the geometry of collider experiments also limits the reach for long lived particles like kaons.

{Importantly, based on the analysis of currently available experimental data}, large deviations from SM predictions remain possible in various NP scenarios \cite{Straub:2013zca,Buras:2015yca,Ishiwata:2015cga,Bordone:2017lsy,Fajfer:2018bfj,Mandal:2019gff,Marzocca:2021miv,Crosas:2022quq,Allwicher:2024ncl}. Furthermore, $K^+ \to \pi^+ \nu \bar{\nu}$ 
decays test interactions of light quarks with third-generation leptons \cite{Bordone:2017lsy,Crosas:2022quq,Davighi:2023iks,Allwicher:2024ncl,Delaunay:2025lhl} and provide insights into the origin of flavour. The missing-mass spectrum, in particular, can constrain the nature of light-quark-neutrino interactions \cite{Deppisch:2020oyx,Gorbahn:2023juq,Buras:2024ewl}. In addition, light NP scenarios such as decays into QCD axions can be tested. If flavour-violating couplings are present \cite{Wilczek:1982rv,Feng:1997tn,Kamenik:2011vy,Bjorkeroth:2018dzu,MartinCamalich:2020dfe}, decays of charged kaons of type $K \to \pi a$ will probe the vectorial axion couplings to $s$ and 
$d$ quarks beyond a scale of $\sim\,10^{12}\, \mathrm{GeV}$.
Furthermore, a measurement of the CP violating $K_L \to \pi^0 a$ will provide an additional constraint on the imaginary part of the axion coupling to $s$ and $d$ quarks.

The precision measurement of this golden charged-kaon decay will deepen our understanding of the fundamental laws of nature and the origin of flavour. It will inform future theoretical activities and guide experimental efforts. Beyond this highlighted decay, additional precision measurements such as $K^+ \to \pi^+ \ell^+ \ell^-$, will stringently test the CKM mechanism, lepton flavour universality, and potential signatures of new light particles and scalar contributions~\cite{Crivellin:2016vjc, DAmbrosio:2024rxv}.

\subsection{Experiments}
Experimental investigations of $K^+ \to \pi^+ \nu \bar{\nu}$ rare decays require dedicated setups, as the typical size and environment prevent multi-purpose collider experiments to select statistically significant clean samples of these processes.
Systematic studies of $K^+$ rare decays date back to the 1990s  and continued in the first years of this century with the experiments E777/E851/E865/E787/E949 at BNL~\cite{Alliegro:1992pp,e865:1999kah,BNL-E949:2009dza}, KLOE at Frascati~\cite{KLOE:2009urs}, 
and NA48/2 at CERN SPS~\cite{NA482:2007ucr}.
Started in 2016, NA62, the successor of NA48/2, is performing the most comprehensive experimental survey of the $K^+$ rare decays to date.
However, the $K^+\to\pi^+\nu\bar\nu$ decay is experimentally challenging, and the goal of the NA62 experiment was to go beyond the mere observation of the BNL experiments, providing a precise measurement that was achieved in \cite{NA62:2024pjp}.

In contrast to BNL, NA62 has adopted a decay-in-flight technique to make use of the high energy protons of the SPS.
High energy has the key advantages to increase the $K^+$ yield per proton, to limit the geometrical acceptance to forward regions and to soften the requirements on low-energy photon rejection.
This last feature is crucial to avoid the otherwise overwhelming $K^+\to\pi^+\pi^0$ background and exploits the Lorentz boost-driven correlation between energy and direction of the two photons from $\pi^0$.
High energy has some drawbacks.
First, the experimental layout has a length of the order of 200\,m that could fit only in the ECN3 cavern at CERN.
Then, kaons cannot be separated in the secondary hadron beam from pions and protons, the amount of which is ten times that of $K^+$.
This second point has led 20 years ago to the design of the Gigatracker, the first silicon pixel detector able to measure time with 100\,ps resolution and to withstand hadron rates of 0.5\,GHz.
The technique of NA62 has proven to be successful.
After reporting evidence of the $K^+\to\pi^+\nu\bar{\nu}$ decay from data collected from 2016 to 2018~\cite{NA62:pnn16,NA62:2020fhy,NA62:2021zjw}, the experiment has recently obtained the first $5\sigma$ observation of this decay by the analysis of the data taken until 2022~\cite{NA62:2024pjp}.
With this result NA62 has measured a $K^+\to\pi^+\nu\bar{\nu}$ branching ratio equal to $13.0^{+3.3}_{-3.0}\times10^{-11}$.
NA62 is scheduled to run until LS3 and foresees to collect a factor of 3 more data than those analyzed so far.
Such a statistics could push the relative uncertainty of the branching ratio down to 15\%, assuming the same measured central value.
While the NA62 technique has shown the advantages of using a high-energy environment to reject background photons and muons, the accidental background from beam-related activities turns out to be predominant and has required some hardware modification from the original design to keep it under control.
Eventually, NA62 has obtained a signal over background ratio, $S/B$, of the order of 1, with the accidental background accounting for $3/4$ of the total.
The effect of this background is a reduction of the pure statistical power of a factor of $\sqrt{1+B/S}$. 
Dedicated data analyses have shown that the signal yield depends non linearly on intensity in a non paralyzable dead time-like manner.
Starting from August 2023, NA62 runs at the optimal intensity condition.
This corresponds to a pace of the order of 15--20 selected $K^+\to\pi^+\nu\bar{\nu}$ candidates per year, where a year is 500K spills and the candidates are counted after background subtraction.

A future experiment able to improve significantly over NA62 could use the same concept, but with upgraded detectors to shift the saturation point of the signal yield towards higher intensities \cite{HIKE-Proposal}.
This experimental approach would offer numerous advantages.
First, the experiment would work in a quasi-linear regime at the intensity of NA62, allowing an increase of signal yield already without rise of intensity.
Second, the signal yield could benefit from an additional intensity increase towards the optimal working point.
Finally, the ten-year experience of NA62 would guarantee robustness of the experimental projections.
Again the $K^+\to\pi^+\nu\bar{\nu}$ decay could lead to developments of cutting-edge detector technologies like in the past. 
An example is a Gigatracker-like detector with 20\,ps time resolution, that would be the first step towards a next generation $K^+\to\pi^+\nu\bar{\nu}$ experiment.
Nowadays this type of technology exists, albeit in R\&D phase~\cite{Addison:2024age}.
Similar timing improvements would be required for the detectors measuring the $\pi^+$, a task provided, for example, by modern photomultipliers.

Significantly higher kaon flux can be realised even without increased primary-proton-beam intensity.
A possible way could be opening the momentum spread of the secondary kaon beam that is limited to 1\% in NA62.
With some modification of the beamline and with detectors able to stand higher rates, a momentum spread of 2\% would be viable and would allow a factor of two increase in intensity.
Preliminary studies suggest that the above would-be experiment could
improve the projected final statistical uncertainty of NA62 in 4
years of data-taking by more than a factor of two.
This precision could allow a $5\sigma$ significance rejection of the SM hypothesis, if NA62 eventually confirmed the presently observed slight excess of $K^+\to\pi^+\nu\bar{\nu}$ candidates.
Besides the main topic of $K^+\to\pi^+\nu\bar{\nu}$, the NA62 experience indicates that an upgraded $K^+$ experiment would significantly increase the sensitivity to a vast suite of NP measurements and searches spanning from lepton-number/flavour violation, to lepton-flavour universality, to feebly interacting particles produced in $K^+$ decays.

\section{FCNC $K_L$ decays}
\subsection{Theory}
Rare semileptonic  $K_L\to\pi^0\nu\bar\nu$ decay is the theoretically most pristine channel to search for physics beyond the SM, and therefore constitutes a flavour probe that is conceptually comparable to EW observables like the $W$ mass and the Higgs self-coupling in probing EW physics. The SM prediction  is $\mathcal{B}(K_L \to \pi^0 \nu \bar{\nu}) = 2.87(7)(2)(23)[2.78(6)(2)(29)]\times 10^{-11}$~~\cite{Anzivino:2023bhp}, based on UTfit~\cite{UTfit:2005ras}[CKMfitter~\cite{Hocker:2001xe}] input, where the  uncertainties are from short-distance, long-distance and parametric effects, respectively.
The dominant uncertainties, currently at the level of $10\%$, are, hence,  parametric in nature, and mainly due to the uncertainty in the determination of the CKM parameters $|V_{cb}|$ and  $\bar\eta$~\cite{Aebischer:2022vky,Anzivino:2023bhp}. Long-distance effects can be determined from $K_{\ell 3}$ decays.

The rare semileptonic $K_L\to\pi^0\ell^+\ell^-$ decay is a unique laboratory for studies of CP violation. Experimental bounds on the branching ratio~\cite{KTEV:2000ngj,KTeV:2003sls} are still about an order of magnitude above the expected SM value of order~$10^{-11}$. 
The SM prediction comprises a calculable direct CP-violating contribution~\cite{Buras:1994qa,Buchalla:2003sj}, an indirect CP-violating term that requires input for $K_S\to\pi^0\gamma^*$ form factors~\cite{NA481:2003cfm,Batley:777240}, and, for the muon channel, a CP-conserving two-photon correction~\cite{Isidori:2004rb}. While, at present, the experimental errors dominate, neither channel having been observed, improved SM predictions will become possible if the $K_S\to \pi^0\ell^+\ell^-$ spectra can be measured in the LHCb $K_S$ programme, to infer improved input for the $K_S\to\pi^0\gamma^*$ form factors.  
Lattice QCD calculations of $K\to\pi\gamma^*$ transitions are also being developed~\cite{RBC:2022ddw,Hodgson:2025iit}, proving particularly challenging for the $K_S$ channel. 
Finally, the $K_L\to\pi^0\gamma^*\gamma^*$ matrix element, required for the calculation of two-photon corrections, could be improved with similar methods as $K_L\to\gamma^*\gamma^*$ for $K_L\to\ell^+\ell^-$~\cite{Hoferichter:2023wiy}.
The latter decay is indeed dominated by the two-photon cut, and due to its CP-conserving nature provides access to couplings that are otherwise only accessible in the $K^+\to\pi^+\nu\bar\nu$ channel~\cite{Isidori:2003ts}. At present, the uncertainty in the SM prediction is still dominant, but significant improvement has been achieved recently by leveraging additional input from $K_L\to\pi^+\pi^-\gamma$ decays and the asymptotic behaviour of $K_L\to\gamma^*\gamma^*$ in a dispersive approach~\cite{Hoferichter:2023wiy}. Further improvements are possible in case of additional data input and/or in combination with calculations in lattice QCD~\cite{Chao:2024vvl}.

\subsection{Experiments}
For $K_L\to\pi^0\nu\bar\nu$ decay, discriminating it from abundant backgrounds from other decays (such as $K_L\to\pi^0\pi^0$, with two lost photons), relies heavily on vetoing the presence of additional final-state particles. 
First efforts to obtain experimental bounds on the branching ratio date back to 1989~\cite{Littenberg:1989ix} and later at KTeV~\cite{E799-IIKTeV:1999iym} triggered  ideas for dedicated experimental efforts with significant progress in conceptualising experimental techniques and developing detector technology~\cite{KAMI:2001cyd,Comfort:2015xx}.
After the first dedicated $K_L\to\pi^0\nu\bar\nu$ experiment 
E391 at KEK achieved an upper limit on the branching ratio of $2.6\times10^{-8}$ at 90\% CL \cite{E391a:2009jdb},
the KOTO experiment at J-PARC, which started in 2013,
obtained an upper limit on the branching ratio of $2.2\times10^{-9}$ \cite{KOTO:2024zbl} in 2024 by using data taken in 2021,
now aiming at reaching an ultimate sensitivity of $10^{-10}$ or better in this decade, falling just short of the  
predicted SM branching ratio, $3\times10^{-11}$.

Beyond 2026, KOTO~II at J-PARC in Japan~\cite{KOTO:2025gvq} represents the only planned facility worldwide dedicated to rare kaon decays, and specifically to $K_L\to\pi^0\nu\bar\nu$. KOTO~II, proposed as a follow-up to KOTO, has been designed to improve on the KOTO sensitivity by two orders of magnitude and observe several tens of SM $K_L\to\pi^0\nu\bar\nu$ signal events with a signal to background ratio $S/B\approx 1$, leading to the discovery of this decay with significance exceeding $5\sigma$ assuming the SM rate. In a scenario of NP providing a 40\% deviation of the branching ratio from the SM prediction, the KOTO~II measurement would exclude the SM prediction at 90\% CL. 

The ongoing search for the $K_L\to\pi^0\nu\bar\nu$ decay at the KOTO experiment at J-PARC has established an upper limit of $2.2\times 10^{-9}$ at 90\% CL on the decay branching ratio, which is not limited by the background~\cite{KOTO:2024zbl}, and is expected to reach a sensitivity below $10^{-10}$ by the end of the decade. It must be emphasised that the recent KOTO result demonstrates the essential validity of the KOTO technique: the background levels in current KOTO data have been reduced to the point where at most only a modest improvement is needed in order to guarantee the background rejection required for KOTO II. The principal improvements needed for KOTO II involve moving from a neutral-beam production angle of 16 degrees to 5 degrees to increase the $K_L$ flux by a factor of 2.6, significantly enlarging the detector to increase the signal acceptance by a factor of 3, and slashing the losses from accidental vetoes by improving the timing performance of the detectors with modern technology. In this regard, KOTO II builds not only on the solid foundations of the KOTO experiment, but also profits from more than two decades of detector development dedicated to the study of $K_L\to\pi^0\nu\bar{\nu}$, initially for experiments like KOPIO and more recently for HIKE \cite{HIKE-Proposal}. The significant contingent of European and American kaon physicists joining the KOTO~II proposal demonstrates both that the interest in the physics programme  is lively and that there is an avenue for the incorporation of detector concepts developed for other experiments to improve the KOTO~II design. 

Relative to the existing KOTO detector, KOTO~II will have significantly enhanced capabilities for charged-particle tracking, allowing for ${\cal B}(K_L\to\pi^0\ell^+\ell^-)$ measurements in a second phase of operation. The expansion of the physics programme  will allow KOTO~II to obtain complete information on FCNC decays in the $K_L$ sector, enhancing the potential for the discovery and understanding of any NP present. All phases of KOTO~II operation will provide opportunities for concomitant measurements of other rare kaon decays, including searches for new particles such as dark photons and axion-like particles.

A European perspective for a measurement of  $K_L\to\pi^0\nu\bar{\nu}$ with sufficient sensitivity for discovery and a measurement of the branching ratio at CERN has also been developed. The projections within this set-up are that 
the decays $K_L\to\pi^0e^+e^-$ and $K_L\to\pi^0\mu^+\mu^-$ could be observed at the $5\sigma$-level for the first time, and branching ratios could be measured to 18\% and 12\%, respectively. The project that also includes an extension of the programme  to measuring ${\cal B}(K_L\to\pi^0\nu\bar\nu)$ to 20\%~\cite{HIKE:2022qra}  is technically feasible provided an initial $K_L$ beam.

\section{$K_L-K_S$ interference}
\subsection{Theory}
A promising avenue for the extraction of short-distance physics parameters from kaon physics is the study of $K_L-K_S$ interference effects in the decay $K\to\mu^+\mu^-$~\cite{Chobanova:2017rkj,DAmbrosio:2017klp,Dery:2021mct,Dery:2022yqc}.
In many ways similarly to the $K_L\to\pi^0\nu\bar\nu$ mode, the CP violating mode in $K\to\mu^+\mu^-$ is an extremely clean probe of BSM physics in the kaon sector, constraining the Wolfenstein parameter $\bar\eta$ with theory uncertainty of ${\cal O}(1\%)$ (non-parametric). In terms of theory uncertainty, the short-distance contribution to $K\to\mu^+\mu^-$ is even cleaner than $K_L\to\pi^0\nu\bar\nu$, since the only hadronic parameter needed for its prediction is the kaon decay constant.

The challenge in $K\to\mu^+\mu^-$ lies in the extraction of the CP-violating, short-distance determined mode.
The two time-integrated decay rates, ${\cal B}(K_L\to\mu^+\mu^-)$ and ${\cal B}(K_S\to\mu^+\mu^-)$, are dominated by long-distance contributions and their theoretical predictions have sizeable uncertainties. 
The extraction of short-distance information would therefore need to involve time-dependent observables, or CP asymmetries. The measurement of time dependence is not ideally suited for any of the current or planned kaon experiments. A hypothetical dedicated setup shows promise in preliminary studies, see Ref.~\cite{Marchevski:2023kab}.
{The study of the CP asymmetry, $A_{\rm CP}(K\to\mu^+\mu^-)$, could be carried out at the LHC~\cite{Dery:TOAPPEAR}, and its feasibility at LHCb is currently under investigation~\cite{KOTOtalk}  (see below for more details on the relevant experimental landscape).}

In comparison with the $K_L\to\pi^0\nu\bar\nu$ and $K_L\to\pi^0\ell^+\ell^-$ modes, the measurement of CP violation in $K\to\mu^+\mu^-$ provides a complementary cross-check of the same SM CKM combination. It is complementary in the experimental methods required, in the theory inputs needed, as well as in its sensitivity to NP. Since it is determined by the same short-distance contribution within the SM as $K_L\to\pi^0\nu\bar\nu$, the ratio, ${\cal B}(K_S\to\mu^+\mu^-)_{\ell=0}/{\cal B}(K_L\to\pi^0\nu\bar\nu)$ is an extremely clean test of the SM, evading $|V_{cb}|$-related uncertainties~\cite{Buras:2021nns}. In terms of NP operators, $K\to\mu^+\mu^-$ is unaffected by tensor operators, and is sensitive to right-handed currents which would leave $K_L\to\pi^0\nu\bar\nu$ unchanged~\cite{Dery:2021vql}.

\subsection{Experiments}

The LHCb experiment at CERN benefits from the huge kaon flux produced by the LHC, of about $10^{15}$ kaons per $\mathrm{fb}^{-1}$ ~\cite{AlvesJunior:2018ldo}, and from its flexible software trigger~\cite{Aaij:2019zbu} that allows selecting low-momentum particles.
The LHCb experiment could access $K_L$/$K_S$ interference by means of a tagged analysis, through the reaction $pp \rightarrow K^0K^- X$~\cite{DAmbrosio:2017klp}. Extra processes can be  $pp \rightarrow K^0\pi^+ X$ or  $pp \rightarrow K^0\Lambda X$, but the charged kaon is by far the dominant one. The tagging power can be as high as $\approx 20\%$ according to preliminary studies using fast simulation~\cite{KOTOtalk, Chobanova:2020vmx}. Such excellent tagging power relies on the low kaon multiplicity in generic $K^0$ events, and would open the room for the study of $K^0$ CP asymmetries at LHCb, especially with the proposed Upgrade-II~\cite{LHCb:2021glh}, which aims at collecting 300 $\mathrm{fb}^{-1}$ of integrated luminosity.
Some estimates on the ultimate precision for $A_{CP}$ in different channels are shown in Table \ref{tab:KSKL}, assuming a tagging power of $20\%$. The ranges of variation in the table reflect different scenarios concerning improvements in background rejection or electron trigger efficiency.
\begin{table}[]
    \centering
    \begin{tabular}{c|c|c|c}
       Decay  & Upgrade-Ia & Upgrade-Ib & Upgrade-II  \\
       \hline
       $K^0\rightarrow\mu^+\mu^-$ & $30\%-80 \%$ & $20\%-55\%$ & $8.5\%-25\%$ \\
        $K^0\rightarrow\pi^0\mu^+\mu^-$ & $\approx 26\%$ & $\approx 18\%$ & $\approx 7.5\%$ \\        
        $K^0\rightarrow\mu^+\mu^-e^+e^-$ & $23\%-33\%$ & $15\%-22\%$ & $6.5\%-9.6\%$ \\
        $K^0\rightarrow\pi^+\pi^-e^+e^-$ & $\approx4.3\%$ & $1.8\%-2.9\%$ & $0.05\%-1.2\%$ \\
    \end{tabular}
    \caption{Expected uncertainties in $A_{CP}$ for several $K^0$ decays, assuming a $20\%$ tagging power is achieved. The ranges shown for $K^0\rightarrow\mu^+\mu^-$ and $K^0\rightarrow\mu^+\mu^-e^+e^-$  reflect different assumptions on background rejection. The ranges shown for $K^0\rightarrow\pi^+\pi^-e^+e^-$ reflect different assumptions on electron trigger improvements at LHCb.}
    \label{tab:KSKL}
\end{table}

The precision achieved in $K^0\rightarrow\mu^+\mu^-$ would allow us to determine the sign of $A_{\gamma\gamma}$ as well as the value of $|\bar\eta|$ from kaon physics (assuming a measurement of ${\cal B}(K_S\to\mu^+\mu^-)$ is available), providing an independent cross-check to the planned KOTO~II analysis.
The precision achieved in $K^0\rightarrow\mu^+\mu^-e^+e^-$ would allow us to test the different BSM scenarios proposed in~\cite{DAmbrosio:2013qmd}.

Studying rare kaon processes in the high-pileup environment of the LHC poses a significant experimental challenge. Achieving unprecedented background suppression and precise characterisation of the experimental conditions is crucial for extracting $A_{CP}$. To this end, novel kaon tagging techniques with high tagging power must be developed, unlocking a wealth of previously inaccessible measurements in strange physics. In particular, a successful strategy for measuring $A_{CP}$ in $K\rightarrow\mu^+\mu^-$ decays would mark the observation of the rarest instance of CP violation to date.

\section{BSM reach and complementarity}
\begin{figure}[t]
\centering
\begin{subfigure}[htb]{0.49\textwidth}
\centering
\includegraphics[width=6.3cm]{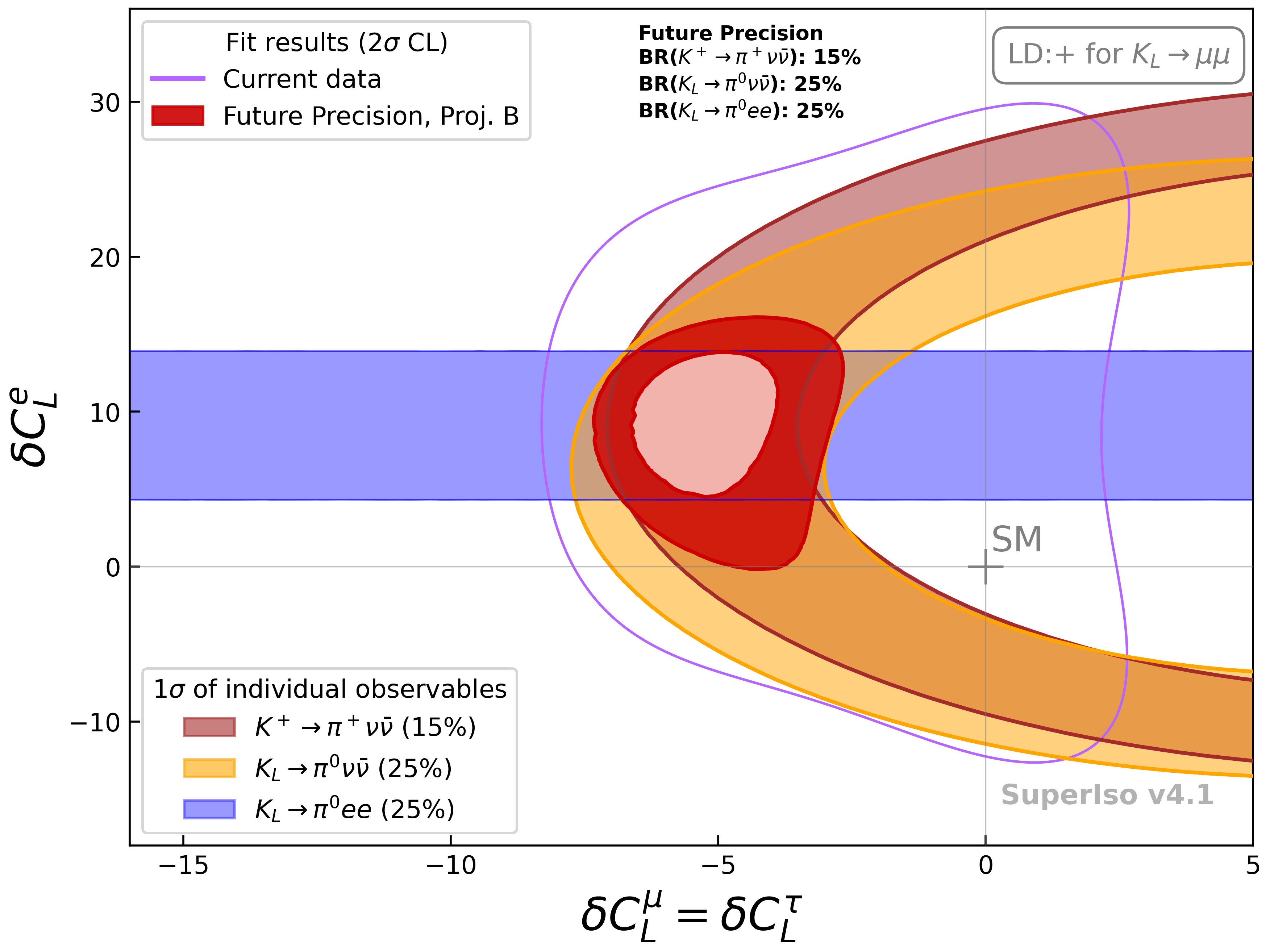}
\caption{}
\label{fig:projectiona}
\end{subfigure}
\begin{subfigure}[htb]{0.49\textwidth}
\centering
\includegraphics[width=6.3cm]{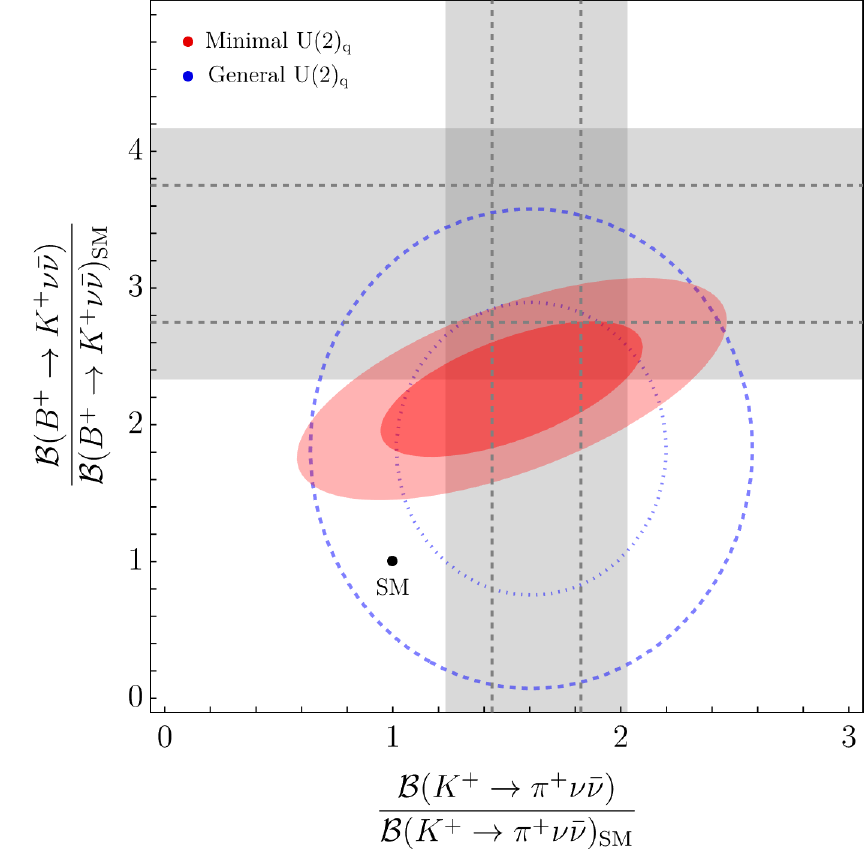}
\caption{}
\label{fig:projectionb}
\end{subfigure}
\caption{The left panel shows the impact of the potential future measurements of $K_L\to\pi^0\nu\bar\nu$, $K^+\to\pi^+\nu\bar\nu$ and $K_L\to\pi^0 e^+ e^-$~\cite{DAmbrosio:2024ewg}. The right panel shows the correlation between di-neutrino modes in the kaon and beauty sector \cite{Allwicher:2024ncl}.
\label{fig:projection}}
\end{figure}
The kaon physics programme  exemplifies the complementarity between different decay modes and sectors. This strengthens its potential and significance in studying 
possible solutions to the flavour puzzle and extensions of the SM that attempt to explain it. 

Figure~\ref{fig:projectiona} illustrates the impact of potential future measurements on NP scenarios in a model-independent framework where NP effects differ for electrons compared to muons and taus. This underscores the strong case for measuring $K_L$ decays with charged leptons in the final state, in addition to the pristine neutrino modes~\cite{DAmbrosio:2024ewg}. The current experimental data sets define the purple region, which represents the allowed parameter space. With the projected sensitivities of $15\%$ from NA62 for $\mathcal{B}(K^+\to\pi^+\nu\bar\nu)$ and 25\% for $\mathcal{B}(K_L\to\pi^0 e^+e^-)$—as well as the KOTO~II projection of 25\% for $\mathcal{B}(K_L\to\pi^0 \nu\bar\nu)$,{ we obtain the pink(red) region at 2(1)$\sigma$}. This highlights that these experimental improvements will significantly reshape the landscape for NP.Z{We also note that the branching ratios of $K^+\to\pi^+\nu\bar\nu$ and $K_L\to\pi^0\nu\bar\nu$ are mutually constrained by an inequality, the Grossman-Nir bound \cite{Grossman:1997sk}, which, however, can be evaded in several BSM scenarios \cite{Kitahara:2019lws,Goudzovski:2022vbt}.}

Figure~\ref{fig:projectionb} demonstrates the complementarity between di-neutrino modes in the kaon ($K^+\to \pi^+\nu\bar\nu$) and beauty ($B^+\to K^+\nu\bar\nu$) sectors within the framework of $U(2)^5$ flavour symmetry for NP couplings~\cite{Barbieri:2011ci}. The current experimental precision (gray bands) shows that the best-fit region (dark and light red for the 1$\sigma$ and 2$\sigma$ contours) is still consistent with this hypothesis. However, future experimental sensitivities (gray dashed lines) could potentially invalidate this scenario~\cite{Allwicher:2024ncl} and can be confronted with a more general scenario in the blue, dashed lines. This underscores the crucial role of kaon physics, in conjunction with the $B$ sector, in distinguishing between different NP scenarios. Interestingly, in this case, the branching ratio of $K_L\to \pi^0\nu\bar\nu$ is highly sensitive to the alignment of NP phases with those of the SM.  
Therefore, combining measurements of $B^+\to K^+\nu\bar\nu$ and $K^+\to \pi^+\nu\bar\nu$ with $K_L\to \pi^0\nu\bar\nu$  
provides crucial insight into CP-violating couplings and reveals potential differences between the beauty and kaon sectors.
{Some NP models also predict correlations between $K_L\to\pi^0\nu\bar\nu$, $K_S\to\mu^+\mu^-$, $K^+\to\pi^+\nu\bar\nu$ and $\epsilon'/\epsilon$~\cite{Buras:2015qea,Aebischer:2023mbz}, which, together with improved lattice calculations for the latter~\cite{RBC:2020kdj} could provide very strong constraints.} We also note that a clean measurement of $K_L\to\pi^0\nu\bar\nu$ decay is of paramount importance for constraining the unitarity triangle from kaon observables alone~\cite{Lunghi:2024sjy}, yielding a precision test of the SM complementary to the $B$-physics programme. 

\section{Conclusions}
The exciting potential of the kaon physics programme, as previously recognised by the European Strategy for Particle Physics, remains highly relevant and complementary to other high-energy physics experiments. Its unique ability to probe NP at the multi-TeV scale distinguishes it as an essential priority for future research efforts, both in Europe and globally. 

Kaon physics offers unparalleled opportunities to characterise NP couplings for light generations and achieve theoretical breakthroughs through the combined use of lattice-QCD, analytical, data-analysis and machine-learning methods. Its precision in rare decays, such as $K^+\to\pi^+\nu\bar\nu$ and $K_L\to\pi^0\nu\bar\nu$, provides clean tests of the SM and sensitivity to NP effects that are inaccessible to current or foreseeable collider experiments. 
Other channels, like $K_L\to\pi^0\ell^+\ell^-$ or $K_L\to\mu^+\mu^-$ offer further enticing opportunities for detailed studies of, e.g., CP violation, while at the same time challenging the theory community to improve their control over long-distance effects.
 {The impact of  kaon experiments is substantial, even in scenarios where no NP is observed, as precise null results would significantly constrain theoretical models.}
  Importantly, no existing or planned experiment duplicates the scope of the kaon programme, ensuring its unique contribution to advancing particle physics. 
  
 The NA62 experiment at CERN has marked a milestone in the history of rare $K$ decays, with the first observation of the theoretically clean rare di-neutrino modes. 
 {Only an upgraded/new version of the NA62 technique, concentrating on $K^+\to \pi^+\nu\bar\nu$ decays, would allow a continued fully comprehensive exploration of $s\to d$ transitions.}
 {The proposed HIKE experiment~\cite{HIKE-Proposal} was set to fulfill this goal, and should serve as 
a foundation for future efforts, at present or forthcoming fixed-target facilities, eg. in the context of a post-LHC hadron facility.}
Together with the KOTO II project in Japan, which focuses on $K_L$ decays, these initiatives are poised to deliver unprecedented precision in rare decay measurements and CP violation studies. 

In conclusion, the kaon programme is indispensable for exploring NP at scales beyond the reach of colliders. It must remain a top priority for the global high-energy-physics community to fully exploit its discovery potential and theoretical implications.

\begin{mybox}
The kaon community requests {to}\\[-6mm]
\begin{itemize}
    \item {protect and amplify the} European kaon-physics programme, exploring opportunities for  \\[-6mm]
    \begin{itemize}
        \item $K^+\to\pi^+\nu\bar\nu$\\[-6mm]
        \item $K_{S,L}\to\mu^+\mu^-$ decay and interference,\\[-6mm]
    \end{itemize}
    \item facilitate and support European contributions for KOTO~II for $K_L\to\pi^0\nu\bar\nu$ and $K_L\to\pi^0\ell^+\ell^-$, { spanning both hardware and analysis development},\\[-6mm]
    \item {maintain the }European leadership on theory computations for kaon physics (phenomenology, dispersion theory, effective theory and lattice QCD, including high-performance computing).
\end{itemize}
\end{mybox}


\clearpage
\normalem
\bibliographystyle{jhep}
\bibliography{biblio}
\end{document}